%% file: main.tex
\documentclass[conference]{IEEEtran} 
\usepackage{cite}
\usepackage[nolist]{acronym}
\usepackage{amsmath,amssymb,amsfonts}
\usepackage{amsmath, mathtools}
\usepackage{algorithm}
\usepackage{algorithmic}
\usepackage{graphicx}
\usepackage{textcomp}
\usepackage{xcolor}
\usepackage{braket}
\usepackage{subfig}
\usepackage{quantikz}
\usepackage{siunitx}
\usepackage{amssymb}   
\usepackage{tabularx}
\def\BibTeX{{\rm B\kern-.05em{\sc i\kern-.025em b}\kern-.08em
    T\kern-.1667em\lower.7ex\hbox{E}\kern-.125emX}}

\title{Grover's Search-Inspired Quantum Reinforcement Learning for Massive MIMO User Scheduling}
\author{\IEEEauthorblockN{Ruining Fan \IEEEauthorrefmark{1}, Xingyu Huang \IEEEauthorrefmark{2}, Mouli Chakraborty \IEEEauthorrefmark{3}, Avishek Nag \IEEEauthorrefmark{4}, Anshu Mukherjee \IEEEauthorrefmark{1}}
\IEEEauthorblockA{\IEEEauthorrefmark{1} School of Electrical and Electronic Engineering,
University College Dublin, Belfield, Dublin 4, Ireland\\
\IEEEauthorrefmark{2}Department of Electrical and Electronic Engineering, Imperial College London, South Kensington, London, UK\\
\IEEEauthorrefmark{3} School of Computer Science and Statistics, Trinity College Dublin, The University of Dublin, Dublin 2, Ireland\\
\IEEEauthorrefmark{4} School of Computer Science,
University College Dublin, Belfield, Dublin 4, Ireland\\
Email: ruining.fan@ucdconnect.ie, anshu.mukherjee@ieee.org
}
}

\begin{document}
\input{acro.tex}
\maketitle

\begin{abstract}
    The efficient user scheduling policy in the \ac{mMIMO} system remains a significant challenge in the field of 5G and \ac{B5G} due to its high computational complexity, scalability, and \ac{CSI} overhead. This paper proposes a novel Grover's search-inspired \ac{QRL} framework for \ac{mMIMO} user scheduling. The QRL agent can explore the exponentially large scheduling space effectively by applying Grover's search to the reinforcement learning process. The model is implemented using our designed quantum-gate-based circuit, which imitates the layered architecture of reinforcement learning, where quantum operations act as policy updates and decision-making units. Moreover, the simulation results demonstrate that the proposed method achieves proper convergence and significantly outperforms classical \ac{CNN} and \ac{QDL} benchmarks by 51\% and 43\%, respectively. 
\end{abstract}
\begin{IEEEkeywords}
    Massive MIMO, User Scheduling, Quantum Reinforcement Learning, Grover’s Search, 5G/\ac{B5G} Networks, Quantum Communication
    \vspace{-0.25cm}
\end{IEEEkeywords}

\section{Introduction} 
Efficient user scheduling remains one of the most highly anticipated algorithmic challenges in \ac{mMIMO} systems. Conventional approaches, while effective in small-scale deployments, suffer from high complexity, limited scalability, and significant CSI overhead in large antenna-user configurations \cite{sabat2025user}. 

Existing works address the problem using various methods, including traditional \ac{ML}-based solutions and some quantum computation-assisted proposals. Due to the constrained computational capabilities of the \ac{BS}s, traditional approaches are restricted to forecasting the \ac{CSI} for upcoming time intervals solely based on the existing dataset. To maintain accuracy, previous studies often require supplementary methods, such as long short-term memory networks \cite{luo2018channel}, for support. Powered by the quantum-based algorithm, the CSI database traversal becomes more achievable, resulting in a higher average sum rate for hybrid quantum communication systems compared to works that use traditional \ac{ML} methods. Some representative papers and their adopted methods are listed in Table 1 for visual comparison. The unique method adopted by our paper is also illustrated, showcasing our originality.
\begin{table}[htbp]
\centering
\caption{Representative Works Methods Comparison}
\begin{tabularx}{\linewidth}{lXXXX}
\hline
\text{Papers} &\cite{yu2021deep} &\cite{bu2019reinforcement} &\cite{huang2025quantum} & \text{\small Our Work}\\
\hline
\textbf{\tiny Deep Learning} & \checkmark &  & \checkmark &  \\
\textbf{\tiny Reinforcement Learning} &  & \checkmark &  & \checkmark \\
\textbf{\tiny Quantum Computing Powered} &  &  & \checkmark & \checkmark \\
\hline
\end{tabularx}
\label{tab:method_comparison}
\vspace{-0.5cm}
\end{table}

Among various quantum algorithms, Grover’s search is particularly suitable for the massive MIMO user scheduling problem, since it can quadratically speed up exploration of unstructured solution spaces. The scheduling task can be viewed as a combinatorial search for optimal user–antenna allocations, where Grover's amplitude amplification could highlight the high-reward policies effectively. Additionally, the oracle can also ensure the exploration–exploitation trade-off in reinforcement learning, making it a natural foundation for our proposed \ac{QRL} scheme.

In this paper, we propose a novel solution by adopting \ac{QRL} for \ac{mMIMO} user scheduling. \ac{QRL} leverages the interplay between reinforcement learning and quantum-inspired search mechanisms to navigate and traverse the exponentially large scheduling space more efficiently. \text{Grover's Search Algorithm} is one of the most essential \ac{QRL}-based searching algorithms, which establishes a quantum search technique that can amplify the probability of finding optimal solutions within an unsorted database quadratically faster than classical algorithms \cite{bulger2003implementing}. Incorporating this principle within the reinforcement learning framework, the \ac{QRL} agent investigates potential scheduling strategies in a manner enhanced by quantum computing. Here, the oracle identifies feasible solutions, and amplitude amplification directs the learning path towards nearly optimal policies. In terms of implementation, we designed a quantum gate-based circuit which imitates the function of the layered structure of a reinforcement learning architecture, where one-to-one quantum operations play the role of decision maker and policy updates. This integration enables our system to reduce the computational burden of \ac{CSI} traversal and to converge faster toward high-throughput scheduling strategies, thereby overcoming the scalability barrier faced by classical methods. Additionally, it also laid the foundation for Grover's search-inspired quantum communication algorithm, which could play a significant role in the fields of 5G and \ac{B5G}. To sum up, the novelty and contribution of this paper can be summarized as follows:
\begin{itemize}
    \item Grover-inspired \ac{QRL} framework for user scheduling: We propose a novel quantum-circuit-based \ac{QRL} method for \ac{mMIMO} user scheduling, where the circuit is integrated with conventional reinforcement learning principles. The design enhances the traversal speed of the computational task, thereby increasing the system's scalability. Simultaneously, he system achieves near-optimal scheduling with higher throughput than classical methods.
    \item Perform evaluation and scalability illustration: The proposed QRL model is compared with conventional \ac{CNN} and \ac{QNN} \cite{huang2025quantum} benchmarks under various scenarios. The scalability of our method is demonstrated by consistently achieving higher average sum rates as the user-antenna configurations are scaled up. The robustness of the system is potentially essential for \ac{B5G} networks.   
\end{itemize}
In the remaining sections, we present the detailed design of the proposed algorithm, followed by numerical analysis and comparative evaluation against existing methods. 

\section{System Model}
In this paper, a single-cell massive MIMO downlink system is considered. The system contains $A$ antennas at the \ac{BS}, which serves $T$ terminal users with a single antenna. For each user, the downlink channel is denoted as $\mathbf{n}_{t} \in \mathbb{C}^{A}$. Although the antenna configurations at the BS have flexibility, we adopt the rectangular configuration in this work: The BS contains $X$ rows of antenna strings and $Y$ antennas are there in each row. Note that the number of users who are served by the BS simultaneously $T \le A=X\cdot Y$. 

\subsection{Received Signals and Ergodic Sum Rate}
In the focused multiuser downlink system, the received signal at each terminal user $t$ can be presented as 
\begin{equation}
    \mathbf{y}_{t} = \mathbf{n}_t\cdot \mathbf{x} + \mathbf{g}_t,
\end{equation}
where $\mathbf{x} = \sum_{t=1}^T\sqrt{p_t}\cdot\mathbf{s}_t$ are the signals transmitted through all the antennas in the BS, with $\mathbb{E}[|\mathbf{s}_t|^2] = 1$ being the data symbols of user $t$. The noise component $\mathbf{g}_t$ follows the pattern of \ac{AWGN} that $\mathbf{g}_t \sim \mathcal{CN}(0, \sigma_t^2)$. For the sake of fairness for every user in the system, we assume each user gets the same amount of energy $ P = \sqrt{p_t} =\sqrt{P_{total}/ T}, \forall{t}$\cite{li2016statistical3-D}. 

Furthermore, the ergodic sum rate can be presented as 
\begin{equation}
    S_{sum} = \sum_{t=1}^{T} S_t,
\end{equation}
where $S_t$ is the so-called ergodic rate, which is deduced by the Shannon capacity and the corresponding concept of \ac{SINR}\cite{Li2019JointScheduling}:
\begin{equation}
    S_t\approx \text{log}_2[1+\mathbb{E}_t(\text{SINR}_t)].
\end{equation}
Since we are dealing with the large-scale Grover's search problem in this paper, which will be discussed in the upcoming section, we can use the approximation form of the ergodic rate for each given user $t$ without affecting the final training outcome of our proposed system. \vspace{-0.25cm}

\subsection{Beam Domain Channel State Information} \vspace{-0.25cm}
Compared to the Rayleigh fading channel conditions that are used widely in large-scale MIMO systems, the Rician fading channel model has a Rician-K factor which balances the channel conditions between pure \ac{LoS} and full scattering \cite{li2016statistical3-D}. It gives more robustness to our proposed algorithm by yielding a flexible feedback under structured channels. Originally, the stacked channel of the system is written as 
\begin{equation}
    \mathbf{N} = [\mathbf{n}_{1},\dots, \mathbf{n}_{u}] \in \mathbb{C}^{\mathbf{A}\times \mathbf{T}},
\end{equation}
which is known as conventional \ac{CSI}. In the considered correlated Rician fading channel, we define the channel vector as
\begin{equation}
    \mathbf{N}_t^r = \tilde{\mathbf{K}}\cdot\mathbf{n}_t^{LoS} +\hat{\mathbf{K}}\cdot \mathbf{n}_t^{NLoS}\sqrt{\mathbf{M}_t},
\end{equation}
where $\tilde{\mathbf{K}}$ and $\hat{\mathbf{K}}$ are the coefficients which are deducted from the Rician K-factor, $\mathbf{n}_t^{LoS}\in \mathbb{C}^{T\times 1}$ denotes the LoS channel vector components, with $\mathbf{n}_t^{NLoS}$ representing the \ac{NLoS} mean-centred components with normalized variance \cite{maaref2008capacityRicianFading}. Moreover, $\mathbf{M}_t \in \mathbb{C}^{T\times T}$ is the channel correlation matrix of the NLoS components.

Now, with the assistance of the Rician fading channel, equation (1) can be represented in a detailed and beam-forming manner \cite{li2016statistical3-D}:
\begin{equation}
\begin{aligned}
    \mathbf{y}_t &= \sum_{t=1}^TP\mathbf{N}_t^r\mathbf{f}_tS_t + \mathbf{g}_t\\
    &=P\mathbf{N}_t^r\mathbf{f}_tS_t + \sum_{x=1,x\ne t}^TP\mathbf{N}_x^r\mathbf{f}_x S_x +\mathbf{g}_t,
\end{aligned}
\end{equation}
note that the $\mathbf{f}_t\in \mathbb{C}^{T\times1}$ is the beamforming vector. Generally, it can be approximated by utilizing the unitary \ac{DFT} $\mathbf{\Omega}_T$ and its conjugate transpose to calculate the diagonal matrix:
\begin{equation}
    \mathbf{D}_t = \mathbf{\Omega}_T^H\mathbb{E}[\|\mathbf{N}_t^r\|_2^2]\mathbf{\Omega}_T, 
\end{equation}
which the $i_t$-th largest diagonal element corresponds to the $j_t$-th in the DFT matrix, which maximizes the energy in a particular direction for an aimed terminal user \cite{sun2015beam}. Therefore, the BS will choose the column as the optimal beamforming vector $\mathbf{f}_t$ for the user accordingly.

\subsection{Proportional Fairness-inspired Problem Formulation}
In this paper, the optimal goal is to schedule users in the targeted mMIMO downlink system so that the entire system can achieve a better sum rate without compromising fairness. Firstly, a scheduling vector is necessary
\begin{equation}
    \mathbf{\theta}(t) = \{\mathbf{\theta}^1(t), \dots, \mathbf{\theta}^T(t)\}\in \{0,1\}^T,
\end{equation}
each scheduling indicator $\mathbf{\theta^t(\cdot)}=1, \forall{t}, t<T$ means the corresponding user is scheduled during the current time slot. From equation (3), the instantaneous achievable rate can be calculated as \cite{huang2025quantum}
\begin{equation}
    \hat{\mathbf{s}}_t(t) = \text{log}_2\left(\frac{\sum_{\gamma=1}^T\mathbf{f}_t\theta^t(t)+1}{\sum_{\gamma=1,\gamma\ne t}^T\mathbf{f}_t\theta^t(t)+1}\right).
\end{equation}
To strike a balance between fairness and channel capacity utilization rate, we choose the \ac{PF} method to ensure fairness for each user.
Thus, the problem can now be written as 
\begin{equation}
    \begin{aligned}
        &\{\mathbf{\theta}^1(t), \dots, \mathbf{\theta}^T(t)\}\ \mathbf{\text{argmax}}\sum_{\gamma=1}^T\frac{\hat{\mathbf{s}}_\gamma(t)}{\bar{\mathbf{s}}_\gamma(t+1)+a},\\
        &\bar{\mathbf{s}}_\gamma(t+1)=
        \begin{cases}
        (1-\omega)\bar{\mathbf{s}}_\gamma(t)+\omega\hat{\mathbf{s}}_\gamma(t),\ \text{if}\ \mathbf{\theta}^t(t) =1,\\
        \bar{\mathbf{s}}_\gamma(t),\ \text{otherwise}.
        \end{cases}
    \end{aligned}
\end{equation}
Note that the time $(t+1)$ represents the long-term achievable ergodic sum rate, and $\omega\in (0,1)$ is the forgetting factor for the historical rate, and $a$ is just a small number that prevents the division by zero problem.

\section{Proposed Quantum Reinforcement Learning Method}
\subsection{Grover's Search Quantum Circuit}
Grover's search is an algorithm that utilizes quantum superposition and parallelism. It can traverse the database quadratically faster than classical computers can in terms of time, assisted by the quantum circuit\cite{giri2017review}. In this section, we provide a detailed description of 
our designed Grover's search quantum circuit. The abstract architecture is illustrated in Fig. 1, with solid lines representing its three main layers. 
\begin{figure}[htbp]
    \centering
    \includegraphics[width=0.8\linewidth]{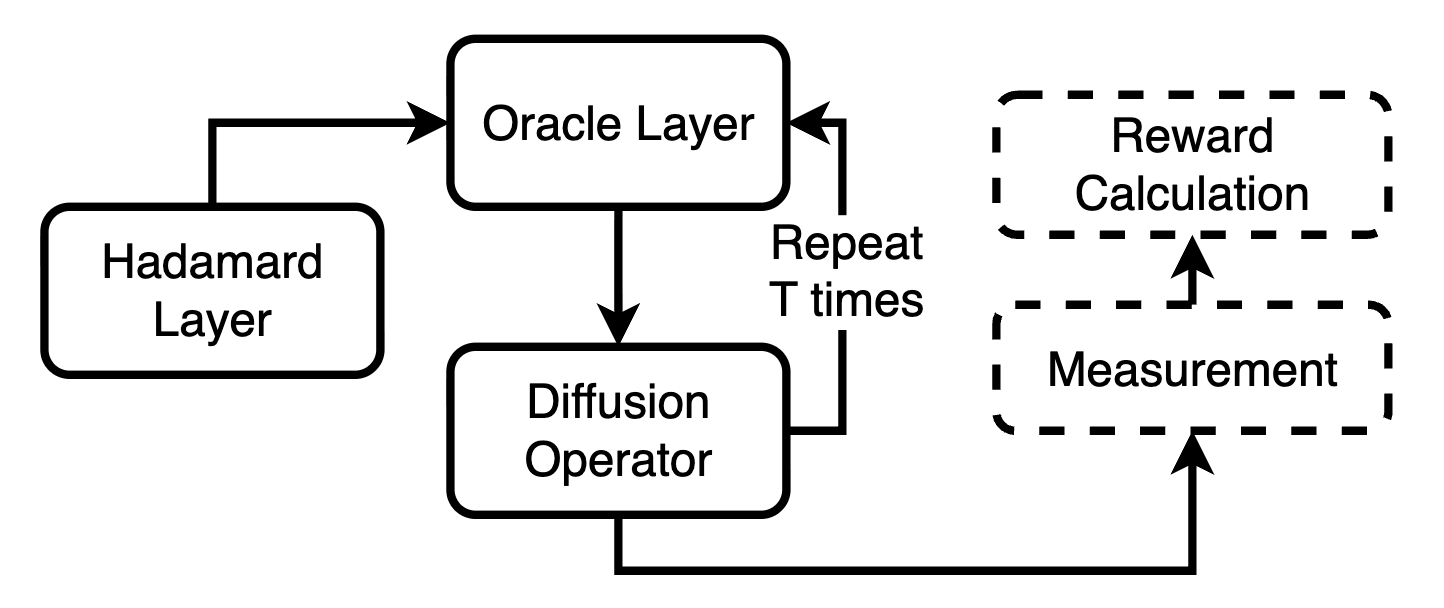}
    \caption{Grover's search-based quantum circuit architecture.}
    \label{fig:placeholder}
\end{figure}

Firstly, the initial superposition layer, also known as the Hadamard layer, is responsible for initializing the state of each input qubit. Hadamard gates create a uniform superposition over all possible states, so that the Grover's agent can search the entire solution space in parallel \cite{shepherd2006role}. Since we have $\mathbf{T}$ users in the targeted system, we require the same number of Hadamard gates and wires accordingly. In the beginning, $\mathbf{N}= \mathbf{T}$ qubits are used to encode all the users,
\begin{equation}
    \ket{\mathbf{\psi}_0} = \ket{0}^{\bigotimes\mathbf{N}},
\end{equation}
which are then fed to the Hadamard gates with the corresponding dimension, and we get
\begin{equation}
    \ket{\mathbf{\psi}_1} = H^{\bigotimes\mathbf{N}}\ket{\mathbf{\psi_0}}=\mathcal{A}\sum_{i=0}^{2^\mathbf{N}-1}\cdot\ket{\theta},
\end{equation}
where $\mathcal{A} = \frac{1}{\sqrt{2^\mathbf{N}}}$, is the uniformed probability amplitude for each superposed states. $\ket{\theta} = \ket{\theta^1\theta^2\dots\theta^{\mathbf{N}}}$ is the computational basis states, converted from the scheduling vector in equation (8).

Furthermore, in the second oracle layer, some positive action in the iteration will be marked, which is done by flipping the phase of the states using Pauli-X pre-processing and a followed multi-controlled Z gate. The marked states are stored in a space $\mathcal{M} \subseteq \{0,1\}^\mathbf{N}$, and the oracle marking action obeys the rule
\begin{equation}
    \mathcal{O}_{\mathcal{M}}\ket{\theta} = \begin{cases}
        -\ket{\theta},\ \theta\in \mathcal{M},\\
        \ket{\theta},\ \theta\notin \mathcal{M}.
    \end{cases}
\end{equation}

Moreover, the diffusion operator layer mirrors and reflects the oracles from the last layer, but centered around the uniform state. This layer mainly relies on a diffusion operator, which is defined as 
\begin{equation}
    \text{Diff} = 2\cdot \ket{U}\bra{U} -I.
\end{equation}
Note that $\ket{U}$ is the uniform state that comes from equation (12), and $I$ denotes the identity matrix. It ensures the core amplification steps in Grover's search, which perform inversion about the mean \cite{giri2017review}.

To summarise, the detailed design of the proposed Grover's search quantum circuit is illustrated in Fig. 2, which uses only 5 qubits for simplicity and can be extended to higher dimensions. 

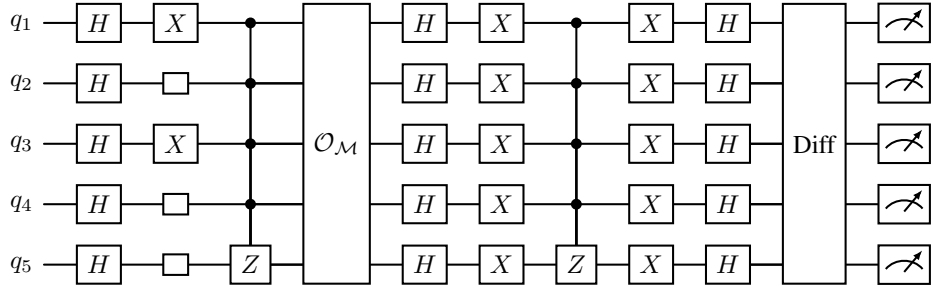
\begin{figure*}[t]
  \centering
\resizebox{0.7\textwidth}{!}{
\begin{quantikz}[row sep=0.3cm, column sep=0.45cm]
\lstick{$q_1$} & \gate{H}
  & \gate{X}               
  & \ctrl{4}               
  & \gate[wires=5, disable auto height]{\mathcal{O}_{\mathcal{M}}}
  & \gate{H}
  & \gate{X}
  & \ctrl{4}
  & \gate{X}
  & \gate{H}
  & \gate[wires=5, disable auto height]{\text{Diff}}
  & \meter{} \\
\lstick{$q_2$} & \gate{H}
  & \gate{\,}              
  & \ctrl{3}
  & \qw
  & \gate{H}
  & \gate{X}
  & \ctrl{3}
  & \gate{X}
  & \gate{H}
  & \qw
  & \meter{} \\
\lstick{$q_3$} & \gate{H}
  & \gate{X}               
  & \ctrl{2}
  & \qw
  & \gate{H}
  & \gate{X}
  & \ctrl{2}
  & \gate{X}
  & \gate{H}
  & \qw
  & \meter{} \\
\lstick{$q_4$} & \gate{H}
  & \gate{\,}              
  & \ctrl{1}
  & \qw
  & \gate{H}
  & \gate{X}
  & \ctrl{1}
  & \gate{X}
  & \gate{H}
  & \qw
  & \meter{} \\
\lstick{$q_5$} & \gate{H}
  & \gate{\,}              
  & \gate{Z}               
  & \qw
  & \gate{H}
  & \gate{X}
  & \gate{Z}               
  & \gate{X}
  & \gate{H}
  & \qw
  & \meter{}
\end{quantikz}
}
  \caption{Grover-based Quantum Circuit (5 qubits).}
  \label{fig:grover-5q}
\end{figure*}

\subsection{Proposed Quantum Reinforcement Learning Algorithm}
In this section, the method for training the aforementioned \ac{QRL} model is discussed. In Algorithm 1, the essential training steps are listed in detail. The model takes a basic Grover vector $\mathbf{G}\{\mathcal{S},\mathcal{K},\mathcal{R}\}$, meaning current state, amplify factor, and reward correspondingly, along with other training parameters such as the one that defines the iteration number, batch size, learning rate, and most importantly, the oracle threshold $\tau$. In terms of the output, it will return the measured and collapsed scheduling vector $\mathbf{\theta}(t)$ for an instantaneous time interval. Initially, the model parameters, including those of the optimizer, are set to their initial state without any bias. Secondly, within each training batch, the scheduling indicators pass through the Hadamard layer, where the vectors are superposed and mapped into the Hilbert space with uniform probability amplitudes. In the second layer, each strategy is evaluated through the oracle, which computes the instantaneous sum rate as the reward of the current epoch. Then, the third step will examine whether the reward is greater than the predefined threshold $\tau$; if so, the third layer, as mentioned in the last section, will take action to mark the action with a high reward via phase inversion. Ultimately, the amplitude amplification step updates the trainable parameters to increase the probability of measuring optimal policies across iterations. Finally, after the current strategy is measured in the vector space, the validation reward is determined based on this, which is logged for monitoring purposes. The same procedure will continue until the model's performance has converged.

\begin{algorithm}[htbp]
\caption{Grover-inspired QRL Training for User Scheduling}
\label{alg:grover_qrl}
\begin{algorithmic}[1]
\REQUIRE Initial QRL parameters $\boldsymbol{G}\{\mathcal{S},\mathcal{K}, \mathcal{R}\}$, learning rate $\eta$, batch size $B$, epochs $N_{\text{epochs}}$, Grover iterations per batch $G$, oracle threshold $\tau$
\ENSURE Trained QRL model with near-optimal scheduling policy $\boldsymbol{\theta(t)}$
\STATE Initialize QRL parameters $\boldsymbol{G}$
\STATE Initialise optimiser
\FOR{epoch $=1$ to $N_{\text{epochs}}$}
    \FOR{each batch of training samples}
        \STATE Prepare uniform superposition of candidate policies $\boldsymbol{\theta} \gets \text{QRL}(\mathbf{A};\boldsymbol{G})$
        \STATE \textbf{Oracle:} evaluate reward:\\ \ \ \ \ \ \ \ \ \ \ $\mathcal{R}\gets\text{calculate\_sum rate}(\mathbf{A},\theta(t),\sigma^2)$
        \IF{$\mathcal{R}\ge\tau$}
        \STATE Mark high-reward policies via phase inversion
        \ENDIF
        \STATE \textbf{Amplitude amplification:} update $\mathcal{K}$ to amplify marked policies
    \ENDFOR
    \STATE Measure $\theta(t)$ to collapse into best scheduling policy
    \STATE Log validation reward
\ENDFOR
\STATE \textbf{return} trained QRL model
\end{algorithmic}
\end{algorithm}
\vspace{-0.25cm}
\section{Results}
To evaluate our proposed QRL model and its training outcomes, we first ensure that the model is properly converged, and then it is compared with the traditional \ac{CNN} benchmark and another \ac{QDL} scheme, as discussed in \cite{huang2025quantum}. 

To begin with, the training curve shown in Fig. 3 clearly demonstrates the proper convergence behavior of the designed Grover-inspired QRL model. The curve starts from an initial average sum rate of 22\text{bps/Hz}, the agent shows a steady increase in performance throughout the 500 training epochs, ultimately reaching close to 32 \text{bps/Hz}. During the early 200 epochs, the higher slope of the curve highlights the model's capability to rapidly improve its policy, while the gradual flattening of the curve beyond epoch 350 indicates that the learning process stabilises and approaches a plateau. In terms of the oscillation, it can be seen that as the epoch number increases, there are fewer fluctuations, confirming that the agent is approaching a near-optimal scheduling policy and also demonstrating the exploration–exploitation trade-off.
\begin{figure}[htbp]
    \centering
    \includegraphics[width=\linewidth]{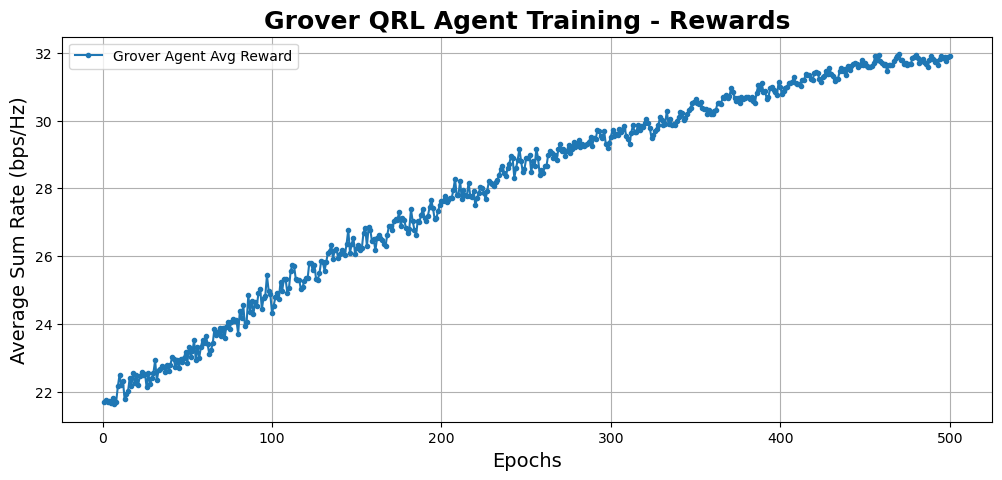}
    \caption{Training convergence of the Grover-inspired QRL agent showing steady reward growth and stabilization.}
    \label{fig:placeholder}\vspace{-0.25cm}
\end{figure}
Now, we move on to the benchmark comparison section. Firstly, we consider the scenario that in a mMIMO system, the antenna number at the BS is fixed, and what would be the performance difference between QRL, QNN and traditional CNN benchmark under different user loads. As illustrated in Fig. 4, the QRL model consistently outperforms both QNN and CNN across all user settings, showing a significant performance gain as the number of users grows. For instance, while all models start around 12 \text{bps/Hz} at $T=2$, the QRL curve rises sharply, reaching nearly 20 \text{bps/Hz} at $T=10$, whereas QNN and CNN achieve only about 17.2 \text{bps/Hz} and 15.8 \text{bps/Hz}, respectively. This result reveals that our proposed QRL framework exhibits better scalability, particularly in the targeted multi-user mMIMO environment, offering not only improved throughput but also robust scalability. 

\begin{figure}[htbp]
    \centering
    \includegraphics[width=0.9\linewidth]{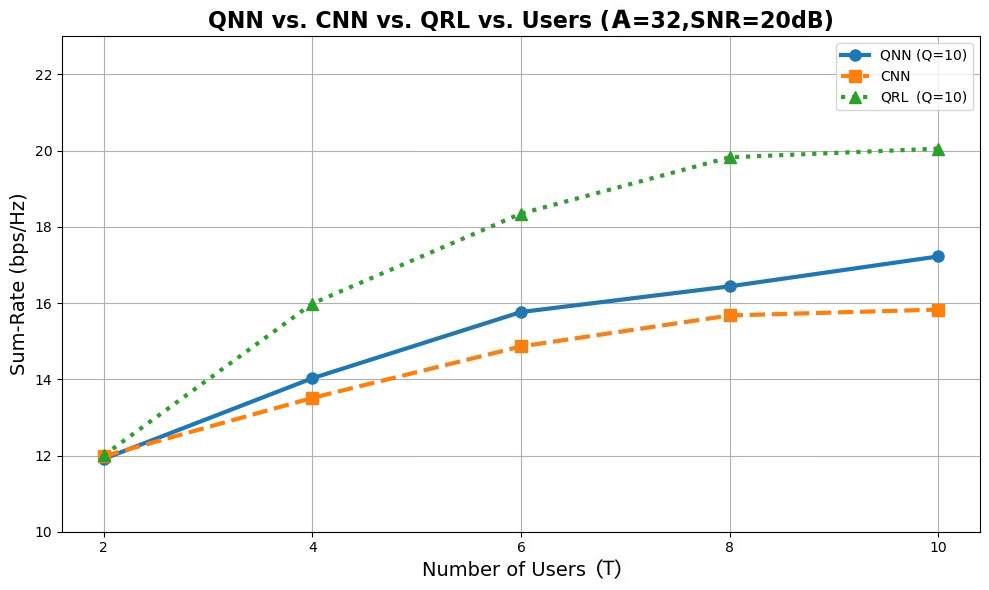}
    \caption{Comparison of sum-rate performance versus number of users for QNN, CNN, and QRL under A=32, \text{SNR}=20 dB, showing the scalability of QRL in user number dimension.}
    \label{fig:placeholder}
\end{figure}

In addition, Fig. 5 presents the average sum rate difference across three methods for 6 users and $\text{SNR}=20$ configurations. The QRL model consistently achieves the highest performance across all antenna settings, starting approxemately 8.2 \text{bps/Hz} at $T=6$ and climbing to nearly 14.7 \text{bps/Hz} at $T=16$. This indicates that QRL leverages the spatial degrees of freedom provided by additional antennas more effectively than its classical and QNN-based counterparts. The QNN baseline performs better than CNN, scaling smoothly with more antennas, yet still falls short of QRL’s throughput gains. There is also a noticeable performance gap between QRL and the benchmarks as the number of antennas increases, illustrating QRL’s scalability in large-antenna regimes. Now we can safely state that the QRL model has scalability in both antenna-wise and user-wise.
\begin{figure}[htbp]
    \centering
    \includegraphics[width=0.9\linewidth]{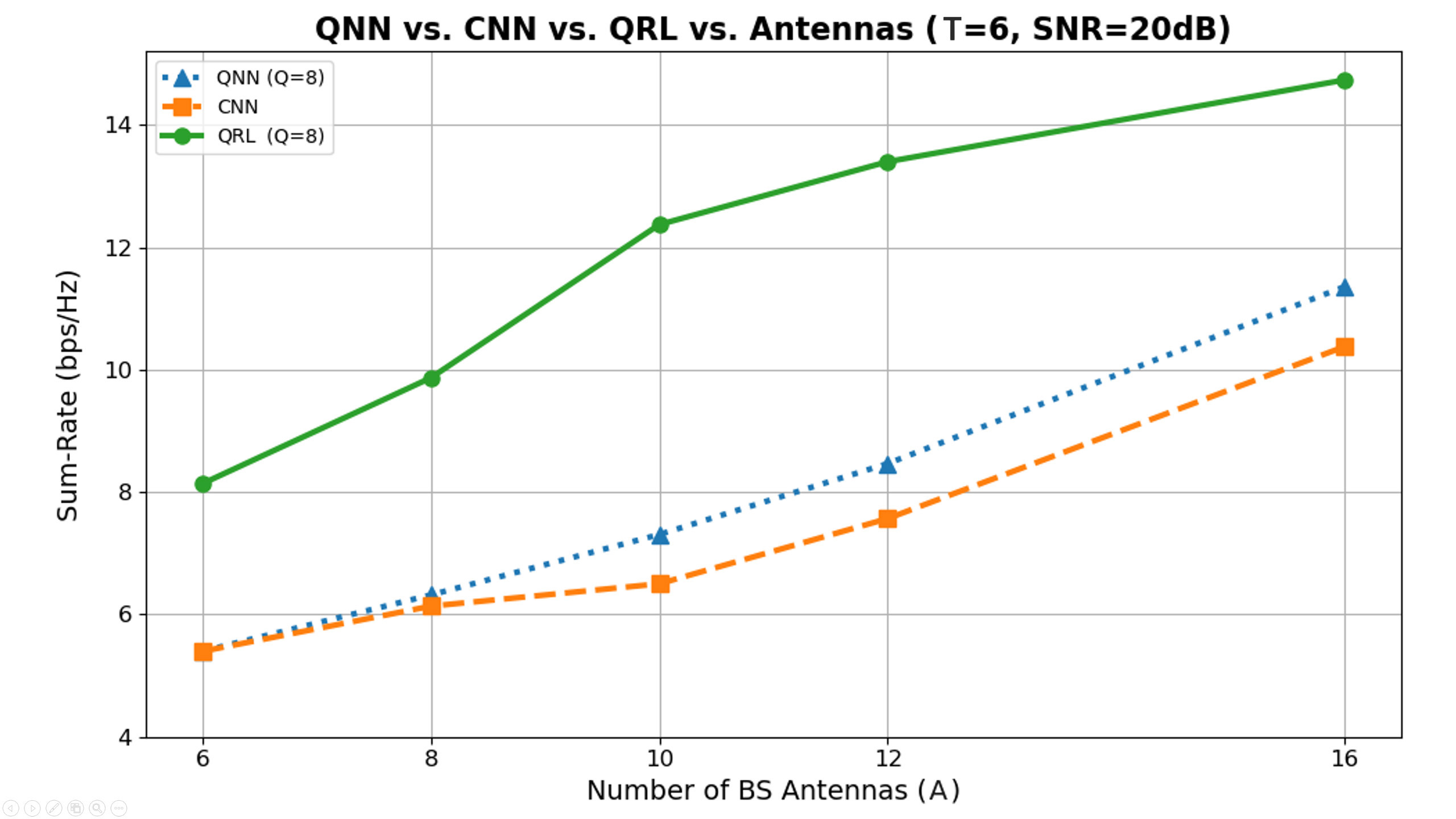}
    \caption{Sum-rate versus number of BS antennas for QNN, CNN, and QRL under $T=6$, \text{SNR}=20 dB, showing QRL’s scalability in antenna number dimension.}
    \label{fig:placeholder} \vspace{-0.25cm}
\end{figure}

Finally, we investigate the performance concerning different $\text{SNR}$ configurations, while the other conditions remain the same. It can be seen that for two different user-antenna sets, the QRL consecutively outperforms its competitors, showing more rapid growth with increasing \text{SNR} in the initial state and saturating at higher sum-rate levels. Note that for the limited configuration, where 6 users and 8 antennas are included, the QRL model has achieved a higher average sum rate level after the $\text{SNR}$ increased by more than 5 \text{dB} compared to the QNN and CNN under a better configuration. It demonstrates the QRL model reacts more sensitively in terms of exploiting high-SNR environments to maximize throughput.
\begin{figure}
    \centering
    \includegraphics[width=\linewidth]{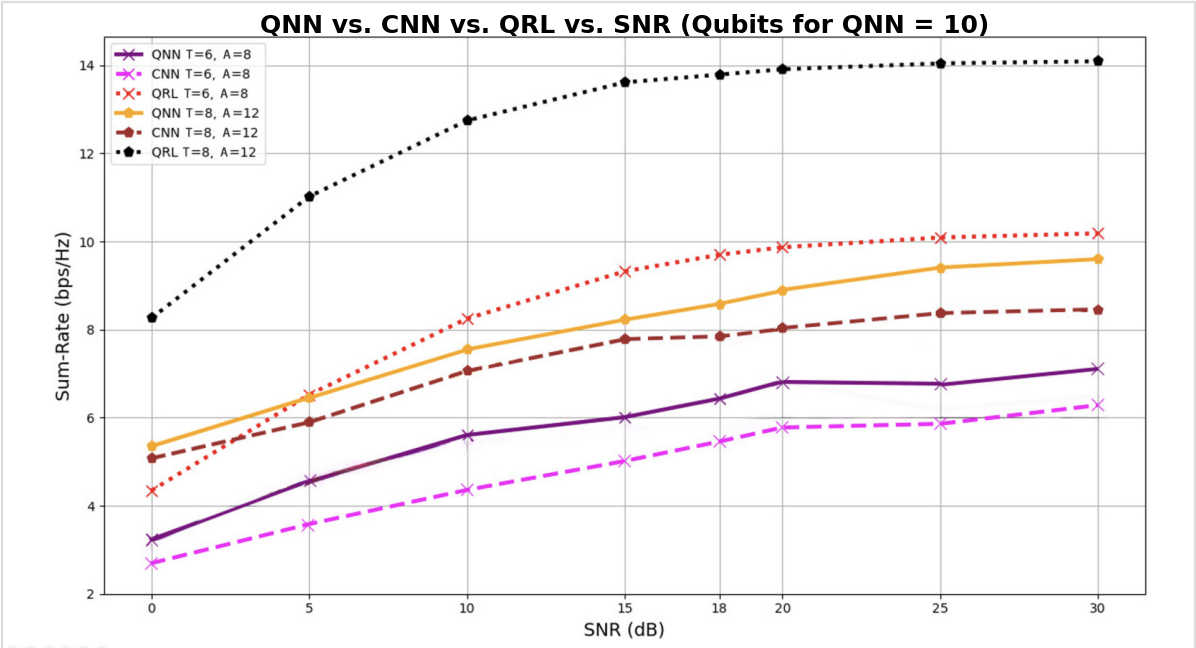}
    \caption{Sum-rate versus SNR for QNN, CNN, and QRL under different user–antenna configurations, highlighting QRL’s sensitivity and scalability in high-SNR environment.}
    \label{fig:placeholder}\vspace{-0.25cm}
\end{figure}

\section{Conclusion}
This paper presents a Grover's search-inspired \ac{QRL} method for user scheduling in \ac{mMIMO} systems. We designed a quantum gate-based circuit that adheres to the principles of reinforcement learning, efficiently exploring the scheduling space and amplifying high-reward policies through amplitude amplification. This design allows the system to overcome the scalability and complexity barriers of classical methods while achieving a higher average throughput. It has been validated by comparison with \ac{QNN} and \ac{CNN} benchmarks across diverse settings, including user and antenna scaling, and varying SNR configurations, demonstrating that the model can not only improve average sum rate but also maintain robustness and adaptability in high-dimensional environments. These findings show the potential of Grover-inspired \ac{QRL} as a promising approach for intelligent resource allocation in 5G/\ac{B5G} networks and future quantum communication networks.

\bibliographystyle{IEEEtran}
\bibliography{reference}
\end{document}

%% file: acro.tex
\begin{acronym}
\acro{B5G}{Beyond 5G}
\acro{TDMA}{Time Division Multiple Access}
\acro{FDMA}{Frequency Division Multiple Access}
\acro{MIMO}{Multiple Input Multiple Output}
\acro{mMIMO}{massive Multiple Input Multiple Output} 
\acro{MU-mMIMO}{Multi-user Massive Multiple Input Multiple Output}
\acro{CSI}{Channel State Information}
\acro{CSIT}{Transmitter Channel State Information}
\acro{BS}{Base Station}
\acro{UT}{User Terminal}
\acro{AI}{Artificial Intelligence}
\acro{ML}{Machine Learning}
\acro{QNN}{Quantum Neural Networks}
\acro{QDL}{Quantum Deep Learning}
\acro{LoS}{Line of Sight}
\acro{NLoS}{Non Line-of-sight}
\acro{QoS}{Quality of Service}
\acro{WMMSE}{Weighted Minimum Mean Square Error}
\acro{DNN}{Deep Neural Networks} 
\acro{RL}{Reinforcement Learning}
\acro{DQN}{Deep Q-networks} 
\acro{PRO}{Policy Optimisation}
\acro{AWGN}{Additive White Gaussian Noise}
\acro{AoD}{Angle-of-departure}
\acro{DFT}{Discrete Fourier Transform}
\acro{ULA}{Uniform Linear Antenna Array}
\acro{NN}{Neural Networks} 
\acro{AP}{Access Point}
\acro{UE}{User Terminal}
\acro{SNR}{Signal-to-noise Ratio}
\acro{SINR}{Signal-to-interference-plus-noise Ratio}
\acro{SLNR}{Signal-to-leakage-plus-noise Ratio}
\acro{GPU}{Graphic Processing Unit}
\acro{PL}{Path Loss}
\acro{ANN}{Artificial Neural Networks} 
\acro{CNN}{Convolutional Neural Networks} 
\acro{GNN}{Graph Neural Networks} 
\acro{dB}{Decibel}
\acro{PF}{Proportional Fairness}
\acro{CNOT}{Controlled-NOT}
\acro{QRL}{Quantum Reinforcement Learning}
\end{acronym}